\documentclass[aps,pra,reprint,amsmath,amssymb,superscriptaddress,showpacs,nobibnotes]{revtex4-1}
\usepackage{graphicx,SIunits}
\usepackage{bm}     
\usepackage{hyperref} 
\usepackage{dsfont}    
\usepackage{color}

\begin{document}


\title{Informationally symmetrical Bell state preparation and measurement}


\author{Yong-Su Kim}
\email{yong-su.kim@kist.re.kr}
\affiliation{Center for Quantum Information, Korea Institute of Science and Technology (KIST), Seoul, 02792, Republic of Korea}
\affiliation{Division of Nano \& Information Technology, KIST School, Korea University of Science and Technology, Seoul 02792, Republic of Korea}

\author{Tanumoy Pramanik}
\affiliation{Center for Quantum Information, Korea Institute of Science and Technology (KIST), Seoul, 02792, Republic of Korea}


\author{Young-Wook Cho}
\affiliation{Center for Quantum Information, Korea Institute of Science and Technology (KIST), Seoul, 02792, Republic of Korea}

\author{Ming Yang}
\email{mingyang@ahu.edu.cn}
\affiliation{School of Physics and Material Science, Anhui University, Hefei, 230601, China}

\author{Sang-Wook Han}
\affiliation{Center for Quantum Information, Korea Institute of Science and Technology (KIST), Seoul, 02792, Republic of Korea}

\author{Sang-Yun Lee}
\affiliation{Center for Quantum Information, Korea Institute of Science and Technology (KIST), Seoul, 02792, Republic of Korea}

\author{Sung Moon}
\affiliation{Center for Quantum Information, Korea Institute of Science and Technology (KIST), Seoul, 02792, Republic of Korea}
\affiliation{Division of Nano \& Information Technology, KIST School, Korea University of Science and Technology, Seoul 02792, Republic of Korea}

\date{\today} 

\begin{abstract}
\noindent  Bell state measurement (BSM) plays crucial roles in photonic quantum information processing. The standard linear optical BSM is based on Hong-Ou-Mandel interference where two photons meet and interfere at a beamsplitter (BS). However, a generalized two-photon interference is not based on photon-photon interaction, but interference between two-photon probability amplitudes. Therefore, it might be possible to implement BSM without interfering photons at a BS. Here, we investigate a linear optical BSM scheme which does not require two photon overlapping at a BS. By unleashing the two photon coexistence condition, it can be symmetrically divided into two parties. The symmetrically dividable property suggests informationally symmetrical BSM between remote parties without a third party. We also present that our BSM scheme can be used for Bell state preparation between remote parties without a third party. Since our BSM scheme can be easily extended to multiple photons, it can be useful for various quantum communication applications.
\end{abstract}

\keywords{Bell state generation, Bell state measurement, Linear optical quantum information processing, Two-photon interference}

\maketitle

\section{Introduction}

Photons are promising physical system for quantum information~\cite{kok07,ralph10,kok16}. Bell state measurement (BSM), which is a projective measurement onto maximally entangled states, plays crucial roles in many photonic quantum information processing applications including quantum teleportation~\cite{bennett93,bouw97}, quantum key distribution~\cite{braunstein12,lo12}, and quantum computation~\cite{gottesman99,klm,gao10}. The linear optical BSM has been widely applied to various quantum information processes due to the simplicity of the experimental implementation~\cite{mattle96, ma12, silva13, choi16}. 


The standard linear optical BSM scheme utilizes Hong-Ou-Mandel (HOM) interference where two photons meet and interfere with each other at a beamsplitter (BS)~\cite{hong87}. Although HOM interference is often understood as a photon bunching phenomenon when two indistinguishable photons meet at a BS, the coexistence of two photons is not essential. Indeed, the physical origin behind HOM interference is not photon-photon interaction at a BS, but interference between two-photon probability amplitudes~\cite{pittman96, kim99, kim03, kim05,kim13,kim14}. This enables two-photon coupling without overlapping two photons at an optical element in time~\cite{wiegner11,megidish13,kim16, Li17}. Therefore, it is fundamentally interesting to investigate whether it is possible to design a new linear optical BSM scheme where two photons do not overlap at a BS. In practice, it might suggest new applications in quantum information processing by unleashing the condition of two photon coexistence from the standard BSM scheme.



In this Letter, we develop a linear optical BSM scheme without photon-photon overlapping at a BS. Our BSM scheme can be also used for preparing Bell states. Unlike the standard linear optical schemes, our Bell state measurement and preparation schemes can be symmetrically divided into two parties. By positioning two parties apart, they inspire informationally symmetrical quantum communication scenarios without a third party. Note that such symmetrical quantum communication cannot be implemented with the standard linear optical Bell state preparation or measurement schemes. We also show that our schemes can be generalized to arbitrary number of $N$-photons, that prepare and analyze $N$-photon Greenberger-Horne-Zeilinger (GHZ) states. Finally, we present proof-of-principle experimental results of our BSM and Bell state preparation schemes.

\section{Linear optical Bell state preparation and measurement}

\subsection{The standard scheme}

\begin{figure}[t]
\includegraphics[width=3.4in]{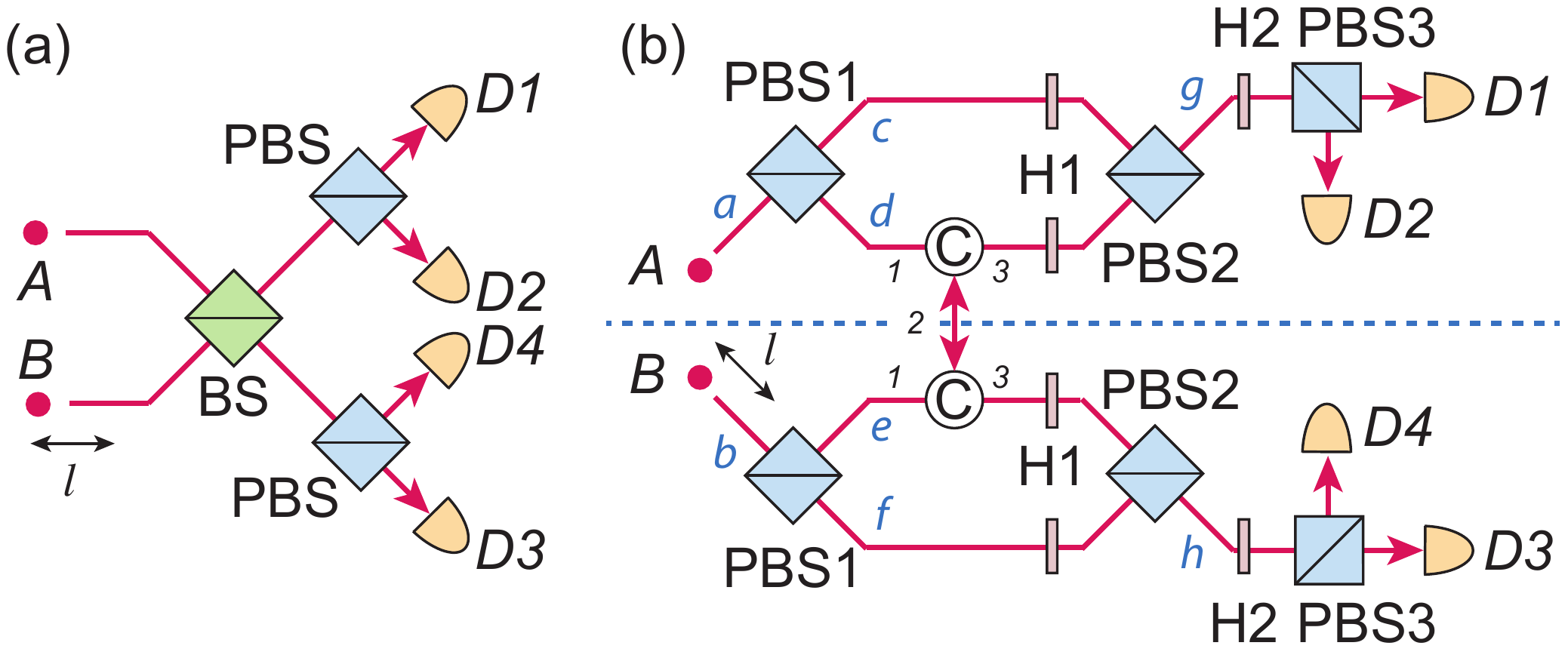}
\caption{(a) The standard linear optical Bell state measurement scheme. Two photons from Alice and Bob meet at a BS. The state $|\psi^+\rangle$ is registered by the coincidences of $D12$ or $D34$, while $|\psi^-\rangle$ corresponds to the coincidences of $D14$ or $D23$. (b) Bell state measurement scheme without interfering particles. Alice and Bob exchange the probability amplitudes of vertical photons using circulators (C) which direct beam $1\rightarrow2$, and $2\rightarrow3$. The state of $|\phi^+\rangle$ is registered by the coincidences of $D13$ or $D24$, while $|\phi^-\rangle$ corresponds to $D14$ or $D23$. $|\psi^\pm\rangle$ cannot be distinguished by this scheme.}
\label{BSM}
\end{figure}

Figure~\ref{BSM}(a) shows the standard linear optical BSM scheme for polarization qubits. The incoming photons from Alice ($A$) and Bob ($B$) meet at a BS, and HOM interference occurs. It is often to scan the optical delay $l$ of one of the photons in order to overlap two photons at a BS. Then, the photons are split by polarizing beamsplitters (PBS) at each output and detected by single-photon detectors ($D1$--$D4$). The state $|\psi^+\rangle=\frac{1}{\sqrt{2}}(|01\rangle+|10\rangle)=\frac{1}{\sqrt{2}}(|HV\rangle+|VH\rangle)$ is registered by the coincidences of $D12$ or $D34$. Here, $|H\rangle$ and $|V\rangle$ denote horizontal and vertical polarization states, respectively, and $Dmn$ is the coincidence between $Dm$ and $Dn$ where $m,n\in\{1,2,3,4\}$. On the other hand, $|\psi^-\rangle=\frac{1}{\sqrt{2}}(|01\rangle-|10\rangle)=\frac{1}{\sqrt{2}}(|HV\rangle-|VH\rangle)$ corresponds to the coincidences of $D14$ or $D23$. The other two Bell states of $|\phi^\pm\rangle=\frac{1}{\sqrt{2}}(|00\rangle\pm|11\rangle)$ cannot be determined by this scheme, so the success probability is $P_s=1/2$. Note that the standard BSM scheme can be used for generating Bell states. For example, if one inputs $|H\rangle$ and $|V\rangle$ states at each input modes, $|\psi^-\rangle$ state is prepared between two outputs of the BS when each output mode occupies a single-photon state.

\subsection{Informationally symmetrical scheme}

Let us introduce our scheme of informationally symmetrical BSM without a third party, see Fig.~\ref{BSM}(b). The probability amplitudes of $|H\rangle$ and $|V\rangle$ of photons $A$ and $B$ are divided by PBS1. Similar to the standard scheme, the optical delay $l$ of one of the photons can be scanned in order to allow interference. While Alice and Bob keep the probability amplitudes of $|H\rangle$, they exchange the probability amplitudes of $|V\rangle$ using circulators (C) which direct beam as $1~\rightarrow~2$, and $2~\rightarrow~3$. After half waveplates (HWP, H1) at $45^{\circ}$, the probability amplitudes interfere at PBS2. Finally, the photons are detected by $D1$--$D4$ after passing through HWP at $22.5^{\circ}$ (H2) and PBS3. 

In order to investigate how the proposed scheme performs BSM, we first consider the initial state of 
\begin{equation}
|\phi^\pm\rangle=\frac{1}{\sqrt{2}}\left(a_H^\dag b_H^\dag \pm a_V^\dag b_V^\dag\right)|0\rangle.
\end{equation}
Here, $a^\dag$ and $b^\dag$ are the creation operators at modes $a$ and $b$, and the subscripts $H$ and $V$ denote horizontal and vertical polarization states, respectively. After the PBS1, which transmits (reflects) horizontal (vertical) polarization state, the state becomes
\begin{equation}
|\phi^\pm\rangle\rightarrow\frac{1}{\sqrt{2}}\left(c_H^\dag f_H^\dag \mp d_V^\dag e_V^\dag\right)|0\rangle.
\end{equation}
Note that the $\pm$ sign changes to $\mp$ due to the relative phase between the transmitting and reflecting modes. The state remains the same after the exchange of probability amplitudes of $d_V^\dag\longleftrightarrow e_V^\dag$ via circulartors. After the half waveplate at $45^\circ$ (H1) and PBS2, the state evolves to
\begin{equation}
|\phi^\pm\rangle\rightarrow\frac{1}{\sqrt{2}}\left(g_V^\dag h_V^\dag \pm g_H^\dag h_H^\dag\right)|0\rangle.
\label{eq3}
\end{equation}
The transformation of a HWP at $22.5^\circ$ (H2) is presented as $g_{H/V}^\dag\rightarrow\frac{1}{\sqrt{2}}(g_H^\dag\pm g_V^\dag)$, and $h_{H/V}^\dag\rightarrow\frac{1}{\sqrt{2}}(h_H^\dag\pm h_V^\dag)$, and thus, the states become
\begin{eqnarray}
|\phi^+\rangle\rightarrow\frac{1}{\sqrt{2}}\left(g_H^\dag h_H^\dag + g_V^\dag h_V^\dag\right)|0\rangle,\nonumber\\
|\phi^-\rangle\rightarrow\frac{1}{\sqrt{2}}\left(g_H^\dag h_V^\dag + g_V^\dag h_H^\dag\right)|0\rangle.
\end{eqnarray}
Therefore, $|\phi^+\rangle$ is registered by the coincidences of $D13$ or $D24$, while $|\phi^-\rangle$ corresponds to $D14$ or $D23$. Note that the successful BSM results are registered by the coincidence between output modes $g$ and $h$. It indicates both outputs $g$ and $h$ are occupied by a single-photon state, and thus, two photons have never met and interacted at an optical element during the propagation for the successful BSM outcomes.

Similarly, before PBS3, the initial state of $|\psi^\pm\rangle=\frac{1}{\sqrt{2}}\left(a_H^\dag b_V^\dag \pm a_V^\dag b_H^\dag\right)|0\rangle$ evolves to 
\begin{equation}
|\psi^\pm\rangle\rightarrow\frac{1}{2\sqrt{2}}\left[\left(g_H^{\dag2}-g_V^{\dag2}\right)\pm\left(h_H^{\dag2}-h_V^{\dag2}\right)\right]|0\rangle.
\end{equation}
It indicates that both $|\psi^+\rangle$ and $|\psi^-\rangle$ states give two photons at one of four possible outcomes $D1-D4$, and cannot be distinguished. Therefore, the overall success probability of BSM is $P_s=1/2$.



Despite the same success probability, there is a significant difference between the standard and our BSM schemes. Let us suppose that Alice and Bob are apart, and each has a single photon. If they want to perform BSM using their photons, in the standard BSM scheme, for example, Alice should send her photon to Bob, then Bob performs BSM. Considering the information flow, this scenario is asymmetric since Alice has to provide her full information (sends her photon) to Bob, whereas Bob does not provide any information to Alice. Since two photons should meet at a BS for the standard BSM scheme, it is impossible to make the scheme symmetric without introducing a third party. The asymmetric nature of the scheme is visualized in Fig.~\ref{BSM}(a) such a way that it cannot be symmetrically divided into the upper and lower parts due to the BS.

On the other hand, our BSM scheme is informationally symmetrical without a third party, i.e., Alice and Bob can perform BSM without breaking the symmetry of the information flow. As presented by the dashed blue line in Fig~.\ref{BSM}(b), it can be symmetrically divided into the upper and lower parties, and they can be apart. Since the successful BSM results are registered by the coincidences of $D13$, $D24$, $D14$, and $D23$, neither Alice nor Bob can retrieve the BSM result alone, but they should cooperate to get the BSM result. 

In cryptologic communication, the information symmetry between two communicating parties, Alice and Bob, is important~\cite{mao}. For instance, if Bob has more information than Alice, it might be possible for him to betray Alice without being noticed. For informationally symmetrical implementation, it is often to have a third party who helps the cryptographic communication between Alice and Bob. However, it requires another assumption that the third party is honest, i.e., she does not cooperate with neither of Alice nor Bob in secret. Therefore, when it is applied to measurement-device-independent (MDI) quantum communication scenarios the informationally symmetrical BSM has a clear advantage over the standard BSM scheme~\cite{braunstein12,lo12,xu14}. It is remarkable that the roles of the third party in standard MDI-QKD scheme are equally shared by Alice and Bob, and thus, the symmetry of the information flow can be maintained.

It is notable that our BSM scheme can be utilized as a scheme to prepare Bell states from separable single photon inputs. Let us consider both Alice and Bob input $|D\rangle=\frac{1}{\sqrt{2}}(|H\rangle+|V\rangle)$. Then, after PBS2, the state evolves to
\begin{equation}
a_D^{\dag}b_D^{\dag}|0\rangle\rightarrow\frac{1}{2}\left(g_H^{\dag}h_H^{\dag}+g_H^{\dag}g_V^{\dag}+h_H^{\dag}h_V^{\dag}+g_V^{\dag}h_V^{\dag}\right)|0\rangle.
\end{equation}
Therefore, if a single-photon state is found at each output of $g$ and $h$, the state becomes a Bell state of
\begin{equation}
a_D^{\dag}b_D^{\dag}|0\rangle\rightarrow\frac{1}{\sqrt{2}}\left(g_H^{\dag}h_H^{\dag}+g_V^{\dag}h_V^{\dag}\right)|0\rangle.
\end{equation}
By positioning Alice and Bob apart, this result suggests informationally symmetrical remote Bell state preparation without a third party. 

\begin{figure}[b]
\includegraphics[width=3.2in]{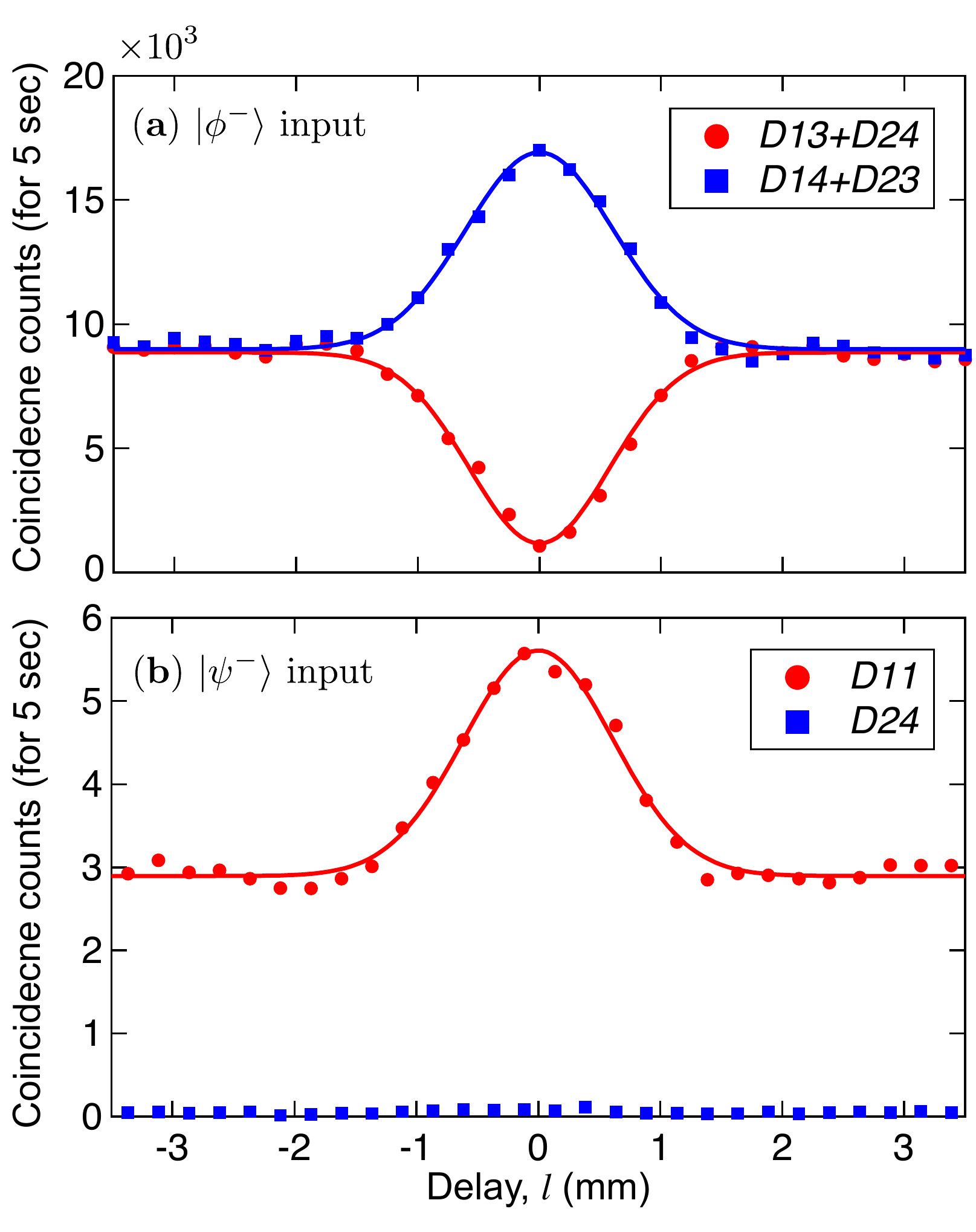}
\caption{Various coincidence counts with respect to scanning the optical delay $l$ for the input states of (a) $|\phi^-\rangle$, and (b) $|\psi^-\rangle$, repectively. When the input state is $|\phi^-\rangle$, one can observe a HOM dip or peak depending on the configurations of the coincidences. The visibility of HOM dip and peak is measured as $V=0.88\pm0.02$ and $0.87\pm0.02$, respectively. For the input state of $|\psi^-\rangle$, the coincidence of $D11$ provides a HOM peak with the visibility of $V=0.94\pm0.02$, while $D24$ shows null outcome.}
\label{interference}
\end{figure}

Similar to BSM, the informationally symmetrical remote Bell state preparation scheme provides advantages to quantum  communication. In many quantum communication protocols, an entangled photon pair is {\it locally} generated by, for example Alice, and distributed between Alice and Bob. In this scenario, Alice has much more information than Bob as she has full control of generating entangled photon pairs. In order to make the scheme symmetric, a third party which generates Bell states and distributes the photon pairs to Alice and Bob is essential. On the other hand, our scheme enables the symmetrical preparation of entangled photon pair between remote Alice and Bob without a third party, and thus, one can implement informationally symmetrical quantum communication without a third party.

Lastly, we remark that our scheme can be generalized to arbitrary number of photons. As shown in the supplement material, by adding the interferometers in parallel, one can generate or perform projective measurement onto $N$-photon GHZ state, $|\phi^\pm\rangle=\frac{1}{\sqrt{2}}\left(|HH\cdots H\rangle\pm|VV\cdots V\rangle\right)$. Therefore, our scheme can be useful for various multi-party quantum communication protocols~\cite{hillery99,fu15}. Note that the feature of symmetrical quantum communication without a third party is still valid.


\section{Experimental results}

The experimental proposal of Fig.~\ref{BSM}(b) requires synchronization and phase stabilization between photons traveling through different optical paths. We note, however, these technical issues can be solved with the current technology even when the optical paths are few hundreds of km~\cite{lucamarini18}. In order to verify our BSM and Bell state preparation schemes, we perform proof-of-principle experiments. Here, we present the experimental results. The experimental details of the photon source and the proof-of-principle experiment can be found in supplement materials.

\begin{figure}[t]
\includegraphics[width=3.2in]{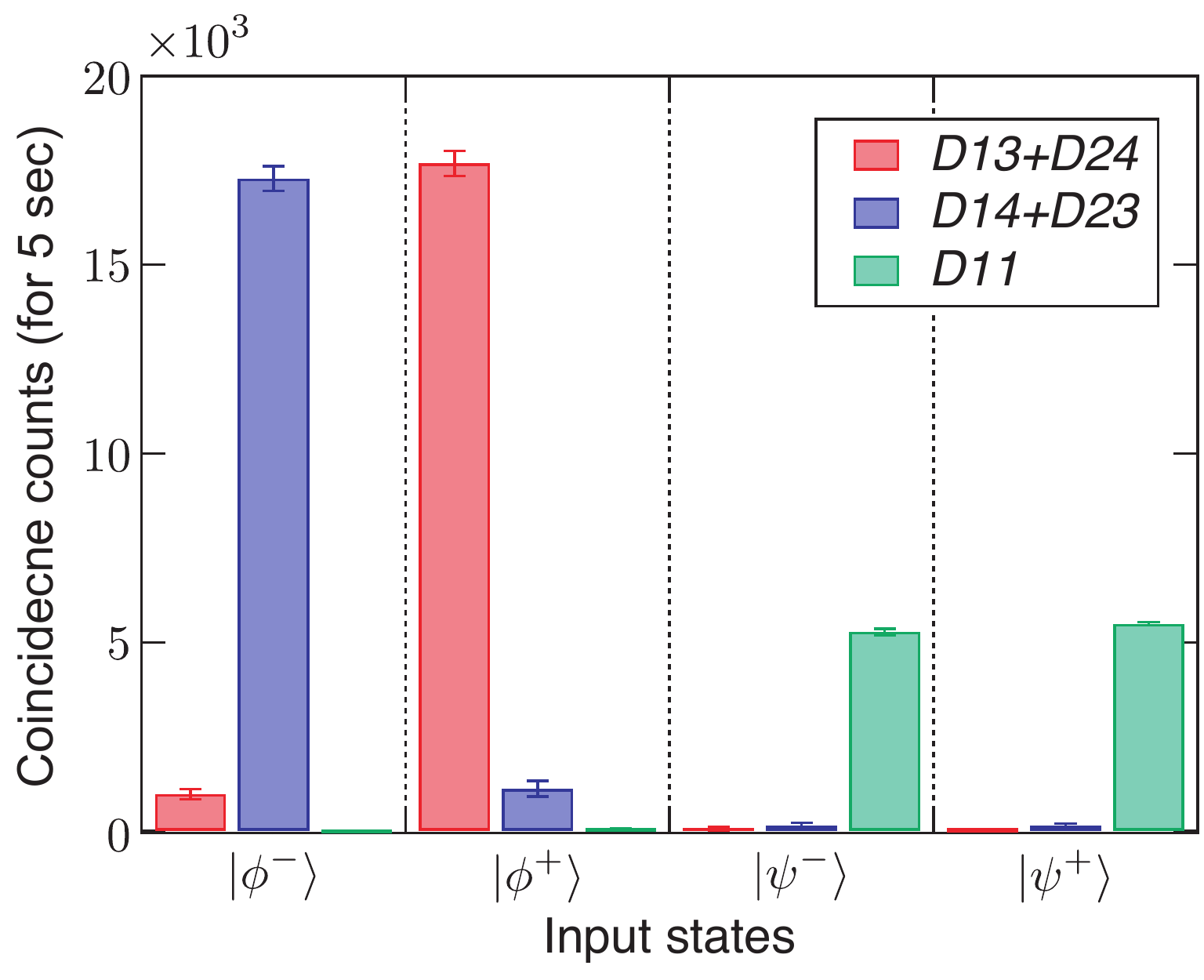}
\caption{Bell state measurement outcomes with respect to various input Bell states. The erroneous detection probability for input $|\phi^-\rangle$ and $|\phi^+\rangle$ states are $Q=5.3\pm0.7\%$ and $5.9\pm1.1\%$, respectively. $|\psi^{\pm}\rangle$ input states do not provide coincidence counts between $D1-D4$, but give two-photon outputs at $D1$ as presented as $D11$. Error bars are experimentally obtained standard deviations.}
\label{Bell_proj}
\end{figure}

Figure~\ref{interference} presents the various coincidences with respect to the scanning optical delay $l$ for the entangled input states of (a) $|\phi^-\rangle$ and (b) $|\psi^-\rangle$, respectively. When the input state is $|\phi^-\rangle$, the coincidences of $D13+D24$ ($D14+D23$) shows a HOM dip (peak) at the optical delay of $l=0$. This result implies that the proposed scheme provides non-classical two-photon interference, and can be useful for BSM. The visibility, which is defined by the relative depth (height) of the dip (peak) to the non-interfering cases, is measured as $V=0.88\pm0.02$ and $0.87\pm0.02$ for the dip and peak cases, respectively. When the input state is $|\psi^-\rangle$, it causes two photon outputs at $D1$, as presented by the HOM peak of $D11$ with the visibility of $V=0.94\pm0.02$. The coincidences of D24 shows null outcome.

In oder to verify the BSM results, we summarize the coincidences depending on the inputs of four Bell states at zero optical delay of $l=0$ in Fig.~\ref{Bell_proj}. Most $|\phi^-\rangle$ states are successfully registered by the coincidences of $D14+D23$ with a few erroneous detections at $D13+D24$. On the other hand, $|\phi^+\rangle$ state provides the successful coincidences of $D13+D24$ while it causes small erroneous detections of $D14+D23$. The input states of $|\psi^\pm\rangle$ do not provide any of above coincidences while they give two photon outputs at the same output port, registered as $D11$. Therefore, we can conclude that the coincidences of $D13+D24$ and $D14+D23$ successfully perform the projective measurement onto $|\phi^+\rangle$ and $|\phi^-\rangle$ states, respectively. Note that the erroneous detection ratios to the overall events of $|\phi^\pm\rangle$ inputs, which corresponds to the quantum bit error rate (QBER) of MDI-QKD, are measured as $Q=5.3\pm0.7\%$ and $5.9\pm1.1\%$ for $|\phi^-\rangle$ and $|\phi^+\rangle$ input states, respectively. These results are well below the lower bound QBER of $11\%$ for secure MDI-QKD, and thus, it is applicable to the MDI quantum communication applications~\cite{lo12}.

\begin{figure}[t]
\includegraphics[width=3.35in]{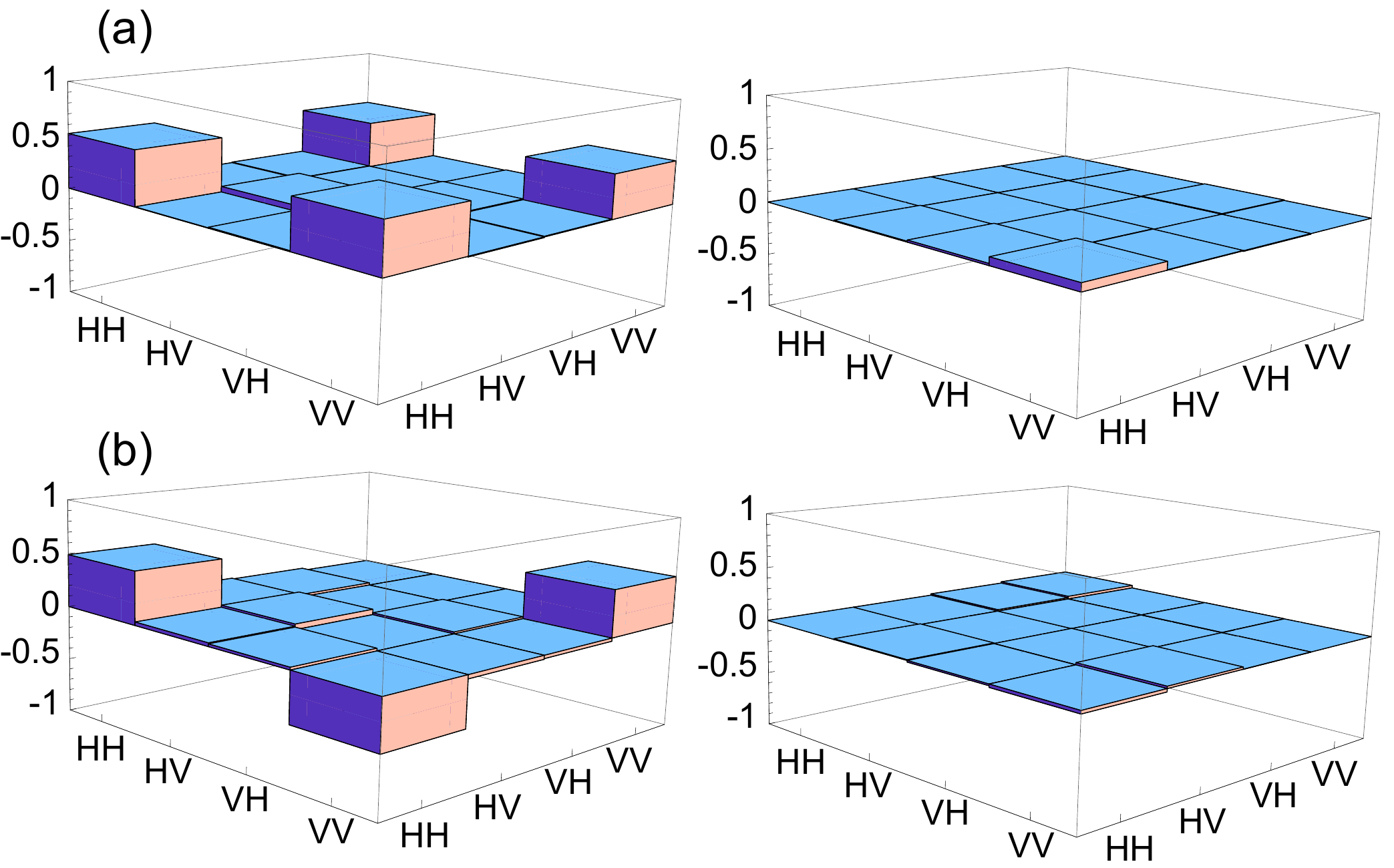}
\caption{Experimentally obtained two-photon states at modes $g$ and $h$ for the separable input states of (a) $|D\rangle_a\otimes|D\rangle_b$, and (b) $|D\rangle_a\otimes|A\rangle_b$, respectively. The left and right figures denote the real and imaginary parts of the density matrix, respectively. (a) The fidelity from the theoretical Bell state of $|\phi^+\rangle$, and concurrence are $F=0.91\pm0.01$, $C=0.85\pm0.03$, respectively. (b) The fidelity from the theoretical Bell state of $|\phi^-\rangle$, and concurrence are $F=0.91\pm0.02$, $C=0.87\pm0.01$, respectively.}
\label{qst}
\end{figure}

In order to verify the ability to generate Bell state with our scheme, we input separable states at mode $a$ and $b$, and perform two-photon quantum state tomography at modes $g$ and $h$~\cite{qst1,qst2}. Figure~\ref{qst} shows  the experimentally obtained two-qubit density matrices for input states of (a) $|D\rangle_a\otimes|D\rangle_b$ and (b) $|D\rangle_a\otimes|A\rangle_b$, where $|A\rangle=\frac{1}{\sqrt{2}}(|H\rangle-|V\rangle)$, respectively. The left (right) figure presents the real (imaginary) part of the density matrix. As predicted by the theoretical investigation, the states are similar to the Bell states of $|\phi^+\rangle$ and $|\phi^-\rangle$, respectively. The fidelities between the theoretical Bell state and the experimentally obtained state are $F=0.91\pm0.02$, and $0.91\pm0.01$, respectively. The estimated concurrence of $C=0.87\pm0.01$ and $0.85\pm0.03$ clearly presents that entangled states can be prepared between two remote parties of Alice and Bob without a third party.

\section{Conclusion}

We present a linear optical Bell state measurement (BSM) scheme which does not require two photon overlapping at a beamsplitter. While the success probability of our BSM is the same as the standard linear optical BSM, i.e., $P_s=1/2$, it has several advantages over the standard BSM scheme. Unlike the standard BSM scheme, our scheme can be symmetrically divided into two parties. The symmetrically dividable property of our scheme suggests informationally symmetrical BSM between remote parties without a third party. We also present that our BSM scheme can be understood as remote Bell state preparation without a third party. The implementation of informationally symmetrical Bell state preparation and measurement enables implementing informationally symmetrical quantum communication which is significant in cryptography. Lastly, we show that our scheme can be generalized to arbitrary number of photons, that prepares or analyzes $N$-photon GHZ states. With the theoretical investigation, we also present proof-of-principle experimental results that prove the effectiveness of Bell state measurement as well as Bell state preparation. Considering fundamental interest and practical applications of remote Bell state preparation and measurement, our scheme will pave a new way to photonic quantum information processing.

\section*{Acknowledgement}
The authors thank M.-S. Kang for useful discussion. This work was supported by the ICT R$\&$D program of MSIP/IITP (B0101-16-1355), and the KIST research program (2E27801). MY is supported by the National Natural Science Foundation of China (NSFC) under Grant No. 11274010.




\begin{thebibliography}{99}

\bibitem{kok07} P. Kok, W. J. Munro, K. Nemoto, T. C. Ralph, J. P. Dowling, and G. J. Milburn, Rev. Mod. Phys. {\bf 79}, 135
(2007).

\bibitem{ralph10} T. C. Ralph and G. J. Pryde, Progress in Optics {\bf 54}, 209 (2010).

\bibitem{kok16} P. Kok, Contemp. Phys., {\bf 57}, 526 (2016).

\bibitem{bennett93} C. H. Bennett, G. Brassard, Claude Cr$\acute{\rm e}$peau, R. Jozsa, A. Peres, and W. K. Wootters, Phys. Rev. Lett. {\bf 70}, 1895 (1993).

\bibitem{bouw97} D. Bouwmeester, J.-W. Pan, K. Mattle, M. Eibl, H. Weinfurter, and A. Zeilinger, Nature {\bf 390}, 575 (1997).

\bibitem{braunstein12} S. L. Braunstein and S. Pirandola, Phys. Rev. Lett. {\bf 108}, 130502 (2012).

\bibitem{lo12} H.-K. Lo, M. Curty, and B. Qi, Phys. Rev. Lett. {\bf 108}, 130503 (2012).

\bibitem{gottesman99} D. Gottesman, and I. L. Chuang, Nature {\bf 402}, 390 (1999).

\bibitem{klm} E. Knill, R. Laflamme, and G. J. Milburn, Nature {\bf 409}, 46 (2001).

\bibitem{gao10} W.-B. Gao, A. M. Goebel, C.-Y. Lu, H.-N. Dai, C. Wagenknecht, Q. Zhang, B. Zhao, C.-Z. Peng, Z.-B. Chen, Y.-A. Chen, and J.-W. Pan, Proc. Natl. Acad. Sci. USA {\bf 107}, 20869 (2010).

\bibitem{mattle96} K. Mattle, H, Weinfurter, P. G. Kwiat, and A. Zeilinger, Phys. Rev. Lett. {\bf 76}, 4656 (1996).

\bibitem{ma12} X.-S. Ma, S. Zotter, J. Kofler, R. Ursin, T. Jennewein, $\breve{\rm C}$. Brukner, and A. Zeilinger, Nature Phys. {\bf 8}, 479 (2012).

\bibitem{silva13} T. F. da Silva, D. Vitoreti, G. B. Xavier, G. C. do Amaral, G. P. Tempor$\tilde{\rm a}$o, and J. P. von der Weid, Phys. Rev. A {\bf 88}, 052303 (2013).

\bibitem{choi16} Y. Choi, O. Kwon, M. Woo, K. Oh, S.-W. Han, Y.-S. Kim, and S. Moon, Phys. Rev. A {\bf 93}, 032319 (2016).




\bibitem{hong87} C. K. Hong, Z. Y. Ou, and L. Mandel, Phys. Rev. Lett. {\bf 59}, 2044 (1987).

\bibitem{pittman96} T. B. Pittman, D. V. Strekalov, A. Migdall, M. H. Rubin, A. V. Sergienko, and Y. H. Shih, Phys. Rev. Lett. {\bf 77}, 1917 (1996).

\bibitem{kim99} Y.-H. Kim, M. V. Chekhova, S. P. Kulik, and Y. Shih, Phys. Rev. A {\bf 60}, R37 (1999).

\bibitem{kim03} Y.-H. Kim, Phys. Lett. A {\bf 315}, 352 (2003).

\bibitem{kim05}  Y.-H. Kim and W. P. Grice, J. Opt. Soc. B {\bf 22}, 493 (2005).

\bibitem{kim13} Y.-S. Kim, O. Slattery, P. S. Kuo, and X. Tang, Phys. Rev. A {\bf 87}, 063843 (2013).

\bibitem{kim14} Y.-S. Kim, O. Slattery, P. S. Kuo, and X. Tang, Opt. Express {\bf 22}, 3611 (2014).

\bibitem{wiegner11} R. Wiegner, C. Thiel, J. von Zanthier, and G. S. Agarwal, Opt. Lett. {\bf 36}, 1512 (2011).

\bibitem{megidish13} E. Megidish, A. Halevy, T. Shacham, T. Dvir, L. Dovrat, and H. S. Eisenberg, Phys. Rev. Lett. {\bf 110}, 210403 (2013).

\bibitem{kim16} H. Kim, S. M. Lee, and H. S. Moon, Sci. Rep. {\bf 6}, 34805 (2016).

\bibitem{Li17} X.-M. Li, M. Yang, N. Paunkovi$\acute{\rm c}$, D.-C. Li, and Z.-L. Cao, Phy. Lett. A {\bf 381}, 3875 (2017).


\bibitem{mao} W. Mao, {\it Modern Cryptography}, (Prentice Hall PTR, New Jersey, 2004).

\bibitem{xu14} P. Xu, X. Yuan, L.-K. Chen, H. Lu, X.-C. Yao, X. Ma, Y.-A. Chen, and J.-W. Pan, Phys. Rev. Lett. {\bf 112}, 140506 (2014).

\bibitem{hillery99} M. Hillery, V. Bu$\breve{{\rm z}}$ek, and A. Berthiaume, Phys. Rev A, {\bf 59}, 1829 (1999).

\bibitem{fu15} Y. Fu, H.-L. Yin, T.-Y. Chen, and Z.-B. Chen, Phys. Rev. Lett. {\bf 114}, 090501 (2015).

\bibitem{lucamarini18} M. Lucamarini, Z. L. Yuan, J. F. Dynes, and A. J. Shields, Nature {\bf 557}, 400 (2018).





\bibitem{qst1} K. Banaszek, G. M. D'Ariano, M. G. A. Paris, and M. F. Sacchi, Phys. Rev. A {\bf 61}, 010304 (1999).

\bibitem{qst2} D. F. V. James, P. G. Kwiat, W. J. Munro, and A. G. White, Phys. Rev. A {\bf 64}, 052312 (2001).





\end{thebibliography}
\end{document}